\documentclass[letter]{aa}  

\usepackage{txfonts}
\usepackage{natbib}
\usepackage{graphicx}
\usepackage{color}
\usepackage{longtable,lscape}
\usepackage{float}
\usepackage{subfigure}

\newcommand{\kmpers}{$\mathrm{km \, s^{-1}}$} % units of velocity

\newcommand{\cmthree}{cm$^{-3}$}

\newcommand{\about}{$\sim$}                       %approx
\newcommand{\expo}[1]{$10^{#1}$}
\newcommand{\texpo}[1]{$\,\times\,10^{#1}$}

\newcommand{\orthowater}{\textit{ortho}-water}     %molecules

\newcommand{\htvao}{H$_2$O}  
\newcommand{\htva}{H$_2$}

\newcommand{\ohtvaoett}{$\rm{H_2O\,(1_{10}-1_{01})}$}
\newcommand{\ohtvaotva}{$\rm{H_2O\,(2_{12}-1_{01})}$}

\newcommand{\amin}{$^{\prime}$}                   %arcus and coordinates
\newcommand{\asec}{$^{\prime \prime}$}
\newcommand{\adeg}{$^{\circ}$}

\newcommand{\atwozero}{$\alpha_{2000}$}
\newcommand{\dtwozero}{$\delta_{2000}$}

\newcommand{\lsun}{$L_{\odot}$}                          %solar and terr. units
\newcommand{\msun}{$M_{\odot}$}

\newcommand{\msunyr}{$M_{\odot} \, {\rm yr}^{-1}$}

\newcommand{\swas}{SWAS} 
\newcommand{\odin}{Odin}
\newcommand{\herschel}{\textit{Herschel}}
\newcommand{\hifi}{{HIFI}}
\newcommand{\pacs}{{PACS}}

\newcommand{\iso}{ISO}

\newcommand{\spitzer}{\textit{Spitzer}}
\newcommand{\vlarhooph}{\mbox{VLA\,1623}}
\newcommand{\mycomment}[1]{} % To hide comments
\newcommand{\renescomment}[1]{} % To hide comments
\newcommand{\perscomment}[1]{} % To hide comments
\newcommand{\myprivatecomment}[1]{} % To hide comments
\titlerunning{Physical properties of outflows from Class~0 sources}
\authorrunning{P. Bjerkeli et al.}

\begin{document}
   \title{Physical properties of outflows\thanks{\herschel\ is an ESA space observatory with science instruments provided by European-led Principal Investigator consortia and with important participation from NASA.}}

   \subtitle{Comparing CO and \htvao\ based parameters in Class 0 sources}

   \author{Bjerkeli, P.
          \inst{1}
          \and
          Liseau, R.\inst{1}
          \and
          Nisini, B.\inst{2}
          \and
          Tafalla, M.\inst{3}
          \and
          Bergman, P.\inst{4}
          \and 
          Melnick, G.\inst{5}
          \and
          Rydbeck, G. \inst{1}
          }

   \institute{Department of Earth and Space Sciences, Chalmers University of Technology, Onsala Space Observatory, 439 92 Onsala, Sweden \\ \email{per.bjerkeli@chalmers.se}
   \and 
   INAF - Osservatorio Astronomico di Roma, Via di Frascati 33, 00040 Monte Porzio Catone, Italy 
   \and Observatorio Astron\'{o}mico Nacional (IGN), Calle Alfonso XII,3. 28014, Madrid, Spain 
   \and Onsala Space Observatory, Chalmers University of Technology, 439 92 Onsala, Sweden 
   \and Harvard-Smithsonian Center for Astrophysics, 60 Garden Street, Cambridge, MA 02138, USA
                \\
                          }
   \date{Received 30 January 2013 / Accepted 7 March 2013}
% \abstract{}{}{}{}{} 
% 5 {} token are mandatory
  \abstract
   {The observed physical properties of outflows from low-mass sources put constraints on possible ejection mechanisms. Historically, these quantities have been derived from CO observations using ground-based observations. 
   %It is reasonable to assume that the momentum is a conserved quantity when outflows are accelerated. 
  It is thus important to investigate whether parameters such as momentum rate (thrust) and mechanical luminosity (power) are the same when different molecular tracers are used.}
   {We aim at determining the outflow momentum, dynamical time-scale, thrust, energy and power using CO and \htvao\ as tracers of outflow activity.}
   {Within the framework of the WISH key program, three molecular outflows from Class~0 sources have been mapped using the \hifi\ instrument aboard \herschel. We use these observations together with previously published \htva\ data to infer the physical properties of the outflows. We compare the physical properties derived here with previous estimates based on CO observations.}
   { Inspection of the spatial distribution of \htvao\ and \htva\ confirms that these molecules are co-spatial. The most prominent emission peaks in \htva\ coincide with strong \htvao\ emission peaks and the estimated widths of the flows when using the two tracers are comparable.}
   {For the momentum rate and the mechanical luminosity, inferred values are independent of which tracer that is used, i.e., the values agree to within a factor of 4 and 3  respectively.}


   \maketitle
%
%________________________________________________________________
\section{Introduction}
The emission from  CO  is a widely used tracer of outflow activity. The lowest rotational transitions emit photons in the millimetre and sub-millimetre part of the electromagnetic spectrum making CO relatively straightforward to observe using ground-based telescopes. Combined with the fact that CO is abundant with respect to molecular hydrogen, this has led to extensive mapping campaigns, covering entire outflows, during the last two decades. Rotational transitions of \htvao\ and \htva\ are, on the contrary, not easily observed from the ground and recent observations have primarily relied on the use of space-based telescopes (e.g. \iso, \swas\ and \odin). However, the limited achievable spatial resolution on these missions hampered the possibility to interpret the data. The situation has improved since the launch of  \spitzer\ \citep{Werner:2004fk} and \herschel\  \citep{Pilbratt:2010kx}. It is now possible to observe  shocked gas, where the water abundance is expected to be enhanced \citep[see e.g.][]{Bergin:1998lr,Kaufman:1996qy}, at a much higher spatial resolution.

Some of the physical properties that can be determined from these observations are of fundamental importance in the understanding of the star formation process.\myprivatecomment{The outflow momentum measure the feedback on turbulence due to star formation and the momentum rate constrains ejection models.} For example, the observed ratio between the momentum inputs of outflows and the luminosities of the central sources sets constraints on the possible ejection mechanisms \citep{Lada:1985kx}. Using CO as a tracer, various parameters of interest have therefore been deduced for several molecular outflows in the past \citep[see e.g.][]{Andre:1990fk,Bachiller:1990lr,Bachiller:2001lr}. It should be noted, however,  that uncertainties in the outflow mass and line-of-sight inclination angle can introduce large errors when estimating the total energy and momentum of the outflowing material.

Within the framework of the \textit{Water In Star-forming regions with Herschel} \citep[WISH,][]{van-Dishoeck:2011lr} key program, three nearby molecular outflows from Class~0 sources were observed using the \hifi\ \citep{de-Graauw:2010uq,Roelfsema:2012lr} instrument aboard \herschel. The mapping observations cover the outflows L1157 \citep{Umemoto:1992yq}, L1448 \citep{Bachiller:1990lr} and \vlarhooph\ \citep{Andre:1990fk,Andre:1993fk}. Observations of the ground-state transition of \orthowater, \ohtvaoett,   at 557~GHz have been discussed, in conjunction with \pacs\ \citep{Poglitsch:2010uq} observations of the \ohtvaotva\ line at 1670~GHz in a series of papers \citep{Nisini:2013qy,Tafalla:2013fk,Bjerkeli:2012fk,Santangelo:2012fk,Vasta:2012uq}. These outflows were also observed with \spitzer\ and discussed in \citet{Neufeld:2009gf}, \citet{Nisini:2010lr}, \citet{Giannini:2011lr} and \citet{Bjerkeli:2012fk}. The analysis of the \htva\ and \htvao\ emission suggests that \htvao\ correlates  to a larger extent with \htva\ than with low-$J$ CO, both in terms of spatial distribution and excitation conditions. The latter molecule is primarily a tracer of the cold, entrained ambient gas while \htva\ and \htvao\ are probing the  shocked gas at elevated temperatures \citep{Tafalla:2013fk,Nisini:2010fkk,Santangelo:2012fk}. \citet{Bjerkeli:2012fk} compare the physical parameters derived from CO and \htvao\ for the \vlarhooph\ outflow. The authors conclude that the force driving the outflow seems to be independent of whether CO or \htvao\ and \htva\ are used as tracers for the outflow. The analysis, however, was based on the observation of a single object and only the north-western outflow lobe of that object. In this letter, we therefore extend this analysis to include also the other outflows that have been mapped with \hifi, i.e. L\,1157 and L\,1448.

We determine the outflow dynamical age, momentum, momentum rate, energy and mechanical luminosity for L1157, L1448 and \vlarhooph. The physical properties derived from the observations of CO(1--0) and CO(2--1) are taken from previously published work. Using the combined spatial and kinematical information obtained from the observations of \htvao\ and \htva\ we then do the same analysis and compare the results.

${ }$
\myprivatecomment{incl in letter? The letter is organized as follows; the observations that our analysis is based on are described in Section~\ref{section:observations}. The inferred results are presented in Section~\ref{section:results} and discussed in Section~\ref{section:discussion}. The main conclusions that can be drawn from this work are presented in Section~\ref{section:conclusions}.}
\begin{figure*}[t]	
\begin{center}
 \rotatebox{0}{\includegraphics[width=14.0cm]{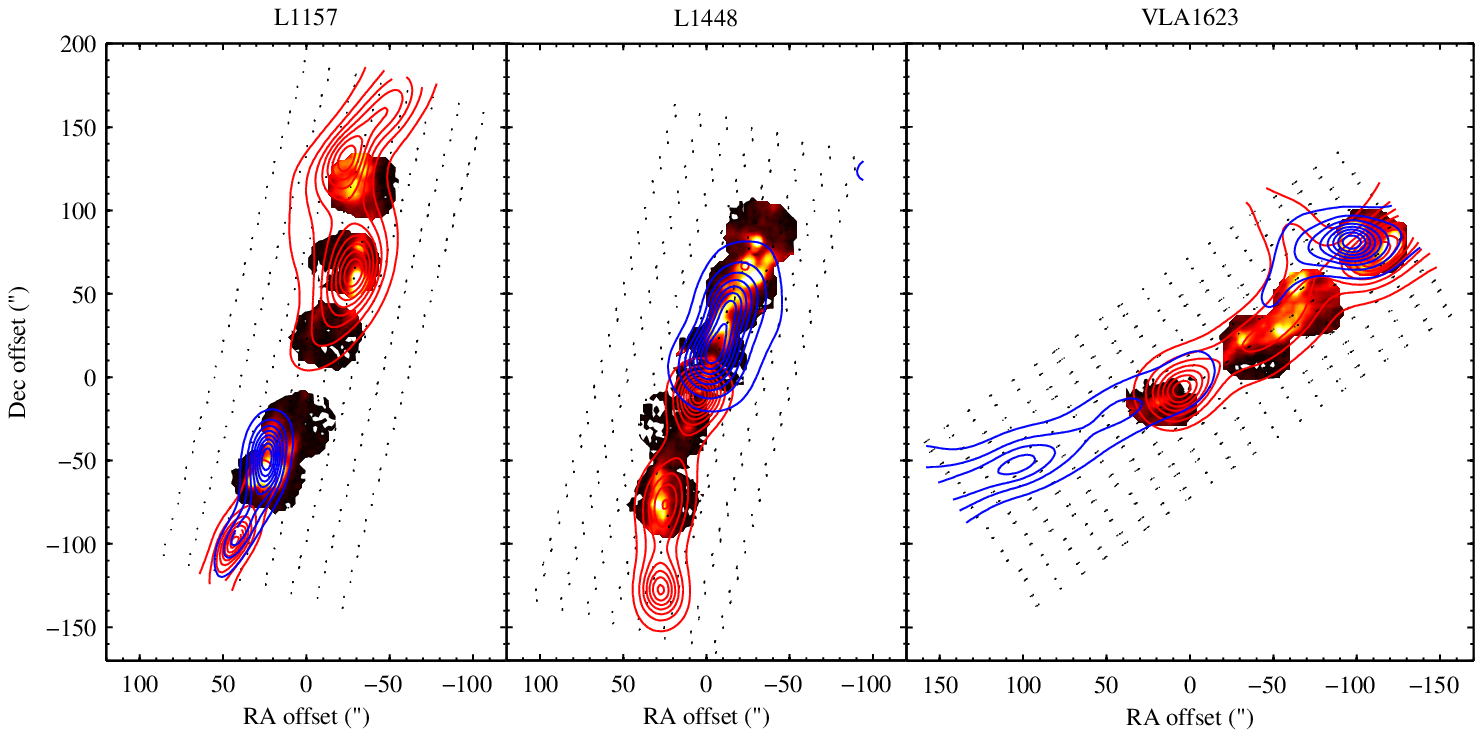}}
     \end{center}
   \vspace{-0.3cm}\caption{Colormaps show the sum of the \htva\ S(0) - S(7) emission (normalized to 1~\texpo{-3}~erg~cm$^{-2}$~s$^{-1}$~sr$^{-1}$)  and contours of the \ohtvaoett\ emission are overlaid. Red contours are from 0.06 to 19 K~\kmpers, 0.09  to 55 K~\kmpers\ and 0.05 to 7.9 K~\kmpers\ for L\,1157, L\,1448 and \vlarhooph, respectively. Blue contours for the same sources are from 0.06  to 69 K~\kmpers, 0.09  to 23 K~\kmpers, and 0.06  to 10 K~\kmpers. Offsets in the maps are with respect to the central sources, with coordinates tabulated in Table~\ref{table:observations}. The black dots indicate the readout positions for the OTF maps.\mycomment{Replace this figure with a new version. It would probably also be a good idea to show the original data (have to think about how though). Also include notes with references and possibly also equations.}}\vspace{-0.2cm}
  \label{figure:maps}
\end{figure*} 
\vspace{-0.4cm}\section{Observations}
\label{section:observations}
The \spitzer\ observations of the purely rotational \htva\ transitions, \htva\ 0 -- 0 S(0) to S(7) were originally presented in \citet{Neufeld:2009gf} and we refer the interested reader to that publication.

The mapping observations of the \ohtvaoett\ emission, from the outflows targeted during the WISH program, have already been published \citep{Nisini:2013qy,Bjerkeli:2012fk} in recent papers, with one exception, viz. L1157. All three maps were obtained with the \hifi\ instrument using the ``On-The-Fly (OTF) Maps with Position-Switch Reference" observing mode. The maps cover the 5\amin\ by 2\amin\ regions closest to the central source of the outflows, which means that a large proportion of the L\,1448 and L\,1157 outflows are covered. In the case of \vlarhooph, only a small part of the southeastern flow is covered by the observations and consequently we exclude this outflow lobe in the analysis. The data were obtained from the Herschel Science Archive (HSA, v8.2.1 of the pipeline) and standard data reduction methods such as baseline removal  and production of maps were done in XS and Matlab. The uncertainty attributed to the calibration is estimated to be \about10\% \citep{Roelfsema:2012lr} and the main beam efficiency at 557~GHz is 0.76 \citep{Olberg:2010kx}. The \herschel\ observations discussed in the present letter are summarised in Table~\ref{table:observations} where the corresponding observation identification numbers are listed.
\begin{table}[t]
  \begin{center}
    \caption{Observations carried out with \hifi.}
  %  \resizebox{\hsize}{!}{
      \label{table:observations}
      \renewcommand{\footnoterule}{} 
      \begin{tabular}{l l l l l}
        \hline
        \hline
        \noalign{\smallskip}
              Target &\atwozero & \dtwozero &$t_{\textrm{\tiny int}}$& Obs. ID \\
      
        & (hr:min:s)& (deg:min:s)&(sec) &  \\
        \noalign{\smallskip}
        \hline
         \noalign{\smallskip}      
        \noalign{\smallskip}
        L\,1448$^{a}$ &03:25:38.40 & +30:44:06.0 & 6468 & 1342203200 \\
            \vlarhooph$^{b}$ & 16:26:26.38 & --24:24:31.0 & 6552 & 1342204010 \\
              L\,1157 & 20:39:06.20 & +68:02:16.0 &6440 & 1342210068 \\

               \noalign{\smallskip}
        \noalign{\smallskip}
  \noalign{\smallskip}	
        \hline
      \end{tabular}
 %   }
  \end{center}
% \tablefoot{
 %\tablefoottext{a}{
 \vspace{-0.2cm}\textbf{Notes.} $^{\textit{(a)}}$\citet{Nisini:2013qy}
 %} \tablefoottext{b}{
$^{\textit{(b)}}$ \citet{Bjerkeli:2012fk}\vspace{-0.5cm}
 %}
\end{table}
\vspace{-0.2cm}\section{Results}
\label{section:results}
For the work presented here, we rely on an estimate of the outflow width and length (when estimating the mass and the dynamical timescale) as measured from the \htvao\ emission, which is why the HPBW of 38\asec\ at 557~GHz turns out to be a problem. We therefore use a Statistical Image Deconvolution (SID) technique to improve the spatial resolution of the \hifi\ data \citep{Rydbeck:2008qy}. 
The deconvolved \hifi\ maps are presented and overlaid on the \spitzer\ \htva\ emission in Fig.~\ref{figure:maps}. 
From the deconvolution of the maps, we conclude that the widths of the flows are overestimated when using the original datasets. However, the apparent width of the northern flow in L\,1157 does not change by a significant amount when the SID method is used, and this part of the flow is likely broader than the southern part. Apart from this single outflow lobe, the outflow lobes have widths of the order \about20\asec. 

The spatial distributions of the \htva\ emission regions are in good agreement with the regions responsible for the \htvao\ emission. The most prominent peaks in the \htva\ emission, from all three outflows, are coincident with strong \htvao\ peaks. However, we note that the \htvao\ peaks towards the central sources of L\,1448 and \vlarhooph\ are not accompanied with strong \htva\ peaks. In the original \htvao\ data, we detect a weak red-shifted component in the blue-shifted part of the L\,1157 outflow. After deconvolution, this emission stands out and the brightness is significantly enhanced. It will, however, not be discussed further here. Note that the \htvao\ line profiles presented in \citet{Nisini:2013qy} and \citet{Bjerkeli:2012fk} trace velocity ranges equivalent to those of CO. Hence, \htvao\ does not seem to be special insofar as it traces gas moving at higher or lower speeds than CO. The maximum detected velocity for both molecules are in close agreement (i.e. within a few \kmpers).
\vspace{-0.2cm}\section{Discussion}
\label{section:discussion}
In this section we compare the physical properties derived from CO observations to those derived from the \htvao\ and \htva\ observations. Although a resemblance (both what regards the spatial distribution and the excitation conditions) between \htvao\ and \htva\ is well established at this point, it should be noted that this analysis relies on the assumption that also the kinematics are the same. The physical properties of the CO outflows have been presented in \citet[][\vlarhooph]{Andre:1990fk}, \citet[][L\,1448]{Bachiller:1990lr} and \citet[][L\,1157]{Bachiller:2001lr}\mycomment{, where the standard CO/\htva\ abundance ratio \expo{-4} were assumed}\mycomment{AndrŽ 1990, Bachiller 1990 and Bachiller 2001 all use X(CO) = \expo{-4}. I have checked this!}. The results presented in these papers are, however, corrected here, using the most recent method for determining these quantities. 
The question of how outflows are accelerated is still a matter of debate. It has been argued, however, that the highly collimated outflows that are seen to emanate from the very youngest objects (i.e. Class~0 sources) are best explained by jet-driven models \citep[see e.g.][]{Masson:1993fk,Arce:2007fk}. We therefore calculate the dynamical time-scale, momentum and kinetic energy ($t_d$, $P$ and $E$ respectively) in the same manner as suggested by \citet[][based on numerical simulations of jet-driven outflows]{Downes:2007lr}. The inclination angles of the flows with respect to the plane of the sky are assumed to be 21\adeg, 15\adeg\ and 9\adeg\ for L\,1448, \vlarhooph\ and L\,1157, respectively \citep{Girart:2001kx,Davis:1999fk,Gueth:1996yq}. It should be noted that uncertainties in the inferred inclination angles
 does not affect our main conclusions since the introduced errors are equivalent for CO and \htvao. The largest relevant contribution to the error budget is instead attributed to the estimated molecular mass $M$ of the outflows. Also, as noted by \citet{Downes:2007lr}, to use the intensity-weighted velocity \textless$\upsilon$\textgreater$_{{lobe}}$ \citep[see e.g.][]{Lada:1996fk}, averaged over the outflow lobe when determining the dynamical time-scale may overestimate the age of the flow (unless the flow is inclined nearly 90\adeg\ with respect to the line of sight). Therefore, and as discussed by the same authors, we instead estimate the outflow dynamical ages from the maximum velocity, i.e., dividing the extent of the flow with the maximum inclination corrected velocity $t_{{d}} = L_{{lobe}} / \upsilon_{{max}}$. The maximum velocity traced by CO is taken from the literature \citep{Andre:1990fk,Bachiller:1990lr,Bachiller:2001lr}. When estimating the momentum of the flow we use the intensity-weighted velocity and we do not apply any inclination correction. \myprivatecomment{According to \citet{Downes:2007lr}, this should be correct to within a factor of two independent on the inclination angle of the outflow.  }For the kinetic energy we use a correction factor of 5, i.e., the uncorrected value is multiplied by this number \citep[][their Fig.~4 \& 5]{Downes:2007lr}. 

The originally published parameters based on CO data are corrected for most recent distance estimates, 250 pc to L\,1157 \citep[][]{Looney:2007lr}, 232 pc to L\,1448 \citep{Hirota:2011fj}, and 120 pc to \vlarhooph\ \citep{Lombardi:2008lr}. The length of the flows are taken to be equal to the maximum extension of the half-power integrated intensity of each lobe. As mentioned earlier, the angular width of all lobes (except the northern flow of L\,1157) are close to 20\asec. The molecular mass of the \htvao\ emitting regions is calculated from the estimated \htva\ column density \citep{Nisini:2010lr,Giannini:2011lr,Bjerkeli:2012fk}. We assume the typical velocity of the wind to be 200~\kmpers, when estimating the mass-loss rate. All inferred quantities are summarised in Table~\ref{table:physparam}. 
\begin{table*}[t!]
  \begin{center}
    \caption{Physical parameters derived from CO and \htvao\  observations.}
    \resizebox{14.2cm}{!}{
      \label{table:physparam}
      \renewcommand{\footnoterule}{} 
      \begin{tabular}{l l l l l l l l l l l}
        \hline
        \hline
        \noalign{\smallskip}
          Source:		&&\multicolumn{2}{c}{L1448} &	\multicolumn{2}{c}{VLA\,1623} & \multicolumn{2}{c}{L\,1157}	\\	
          	 		&&\multicolumn{2}{l}{$\overline{\rule{3.0cm}{0pt}}$} & \multicolumn{2}{l}{$\overline{\rule{3.0cm}{0pt}}$}    & \multicolumn{2}{l}{$\overline{\rule{3.0cm}{0pt}}$} \\ [-3pt]					
							   \noalign{\smallskip}
							   \noalign{\smallskip}		
Molecular tracer:	&  &	CO 	&\htvao\  &	CO 	&\htvao\ &	CO &	\htvao\ 	\\
\noalign{\smallskip}
\noalign{\smallskip}
\underline{Red-shifted lobe}	\\	
\noalign{\smallskip}	
Total mass (\msun):	&		&	1.6\texpo{-2}	&	6.3\texpo{-3}	&	1.1\texpo{-2}	&	1.6\texpo{-3}	&	8.4\texpo{-2}	&	1.3\texpo{-2}	\\
$L_{\rm{lobe}}$ (pc)	&		&	0.17	&	0.17	&	0.09	&	0.09	&	0.21	&	0.22	\\
\textless$\upsilon$\textgreater$_{\rm{lobe}}$ (\kmpers):	&		&	75	&	64	&	27	&	60	&	75	&	152	\\
$\upsilon_{\rm{max}}$ (\kmpers):	&		&	195	&	218	&	116	&	116	&	143	&	192	\\
$t_{\rm{dyn}}$ (yr):	&		&	8.7\texpo{2}	&	7.7\texpo{2}	&	7.2\texpo{2}	&	7.6\texpo{2}	&	1.4\texpo{3}	&	1.1\texpo{3}	\\
Momentum (\msun\ km s$^{-1}$):	&		&	0.42	&	0.14	&	0.08	&	0.03	&	0.98	&	0.30	\\
Kinetic energy (erg):	&		&	5.6\texpo{44}	&	1.6\texpo{44}	&	2.7\texpo{43}	&	2.0\texpo{43}	&	5.7\texpo{44}	&	3.5\texpo{44}	\\
Momentum rate (\msun\ km s$^{-1}$ yr$^{-1}$):	&		&	4.8\texpo{-4}	&	1.9\texpo{-4}	&	1.1\texpo{-4}	&	3.3\texpo{-5}	&	7.0\texpo{-4}	&	2.7\texpo{-4}	\\
Mechanical luminosity (\lsun):	&		&	5.3\texpo{0}	&	1.8\texpo{0}	&	3.1\texpo{-1}	&	2.1\texpo{-1}	&	3.4\texpo{0}	&	2.6\texpo{0}	\\
Wind mass-loss rate (\msun\ yr$^{-1}$):	&		&	2.4\texpo{-6}	&	9.4\texpo{-7}	&	5.5\texpo{-7}	&	1.7\texpo{-7}	&	3.5\texpo{-6}	&	1.3\texpo{-6}	\\
\noalign{\smallskip}
\noalign{\smallskip}
\underline{Blue-shifted lobe}	\\	
\noalign{\smallskip}
Total mass (\msun):	&		&	2.0\texpo{-2}	&	2.8\texpo{-3}	&		&		&	1.2\texpo{-1}	&	9.4\texpo{-3}	\\
$L_{\rm{lobe}}$ (pc)	&		&	0.16	&	0.08	&		&		&	0.12	&	0.12	\\
\textless$\upsilon$\textgreater$_{\rm{lobe}}$ (\kmpers):	&		&	--75	&	--94	&		&		&	--29	&	--49	\\
$\upsilon_{\rm{max}}$ (\kmpers):	&		&	--195	&	--195	&		&		&	--68	&	--128	\\
$t_{\rm{dyn}}$ (yr):	&		&	7.8\texpo{2}	&	3.8\texpo{2}	&		&		&	1.8\texpo{3}	&	9.4\texpo{2}	\\
Momentum (\msun\ km s$^{-1}$):	&		&	0.53	&	0.09	&		&		&	0.54	&	0.07	\\
Kinetic energy (erg):	&		&	7.1\texpo{44}	&	1.6\texpo{44}	&		&		&	1.2\texpo{44}	&	2.7\texpo{43}	\\
Momentum rate (\msun\ km s$^{-1}$ yr$^{-1}$):	&		&	6.7\texpo{-4}	&	2.5\texpo{-4}	&		&		&	3.0\texpo{-4}	&	7.6\texpo{-5}	\\
Mechanical luminosity (\lsun):	&		&	7.5\texpo{0}	&	3.4\texpo{0}	&		&		&	5.8\texpo{-1}	&	2.4\texpo{-1}	\\
Wind mass-loss rate (\msun\ yr$^{-1}$):	&		&	3.4\texpo{-6}	&	1.2\texpo{-6}	&		&		&	1.5\texpo{-6}	&	3.8\texpo{-7}	\\
  \noalign{\smallskip}
  \noalign{\smallskip}	
        \hline
      \end{tabular}
   }
  \end{center}
% \tablefoot{
 %\tablefoottext{a}{
\vspace{-0.1cm} \textbf{Notes.} Inclination angles with respect to the plane of the sky are $21^{\circ}$, $15^{\circ}$, and $9^{\circ}$ for L\,1448, VLA\,1623 and L\,1157, respectively.
 %} 
  %\tablefoottext{b}{
  The velocity of the wind is assumed to be 200~\kmpers.
  %} 
% }
\end{table*}

The highest detected velocity of the \htvao\ emission line profiles is consistent with previous estimates from CO observations (see Tab.~\ref{table:physparam}).
The dynamical ages of the  outflows are of the order \expo{3} years and this confirms that these sources are very young. They may therefore still be in the formation stage where material is accreted onto the proto-stellar condensation. 
The dynamical age can be compared to the shock propagation time-scale (the time the gas is in a shocked state) for various types of plane-parallel shock models \citep{Bergin:1998lr,Flower:2010fkq}. For C-type shocks, time-scales are of the order \expo{3} years for a pre-shock density of \expo{4}~\cmthree\ and even shorter than this for J-type shocks or when pre-shock densities are higher. The dynamical ages can therefore be longer than the time-scales for shock propagation. For the outflows discussed here, the water emitting regions trace gas at densities of \about \expo{6}~\cmthree\ and temperatures higher than 100~K . Under these conditions, the time-scale for freeze-out onto grains should be longer than the dynamical time-scale \citep{Bergin:1998lr,Hollenbach:2009lr}\myprivatecomment{For gas densities of the order \texpo{6} we have freez-out time-scales of the order \texpo{4} years (Bergin). According to Hollenbach, water does not freeze out unless it is colder than 100 K, and that is clearly not the case.}. Thus, it is possible that once the \htvao\ abundance is enhanced due to the presence of a shock it will not be significantly reduced in the post-shock region. This is in agreement with the fact that the \htvao\ emission does not trace gas at particularly high velocities. 
\begin{figure}[t]	
 \rotatebox{0}{\includegraphics[width=7.8cm]{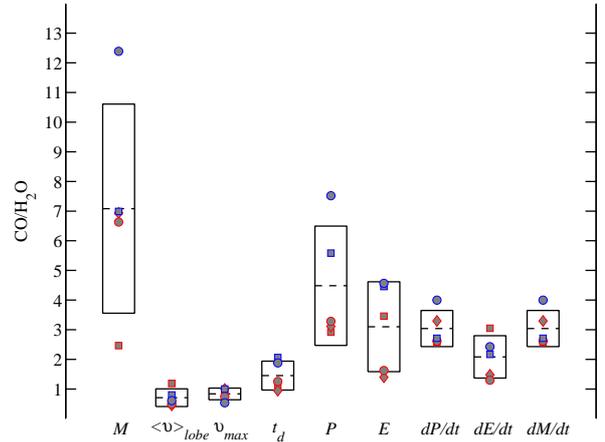}} 
  \caption{
   Ratio between parameters inferred from CO and \htvao\ observations. The markers indicate the ratios for L\,1448 (squares), \vlarhooph\ (diamonds) and L\,1157 (circles). Red and blue colors are for the red and blue outflow lobes, respectively. Note that a few of the symbols are superpositioned. Dashed lines show the mean and boxes indicate the standard deviation.
  }\vspace{-0.1cm}
  \label{figure:ratio}
\end{figure} 
The thrust $dP/dt$ (momentum rate) and power $dE/dt$ (mechanical luminosity) derived from \htvao\ and \htva\ are in close agreement with previous estimates based on CO observations. However, although the deviations are small, the thrust is on average 3 times higher and the power is 2 times higher when deduced from CO observations (see Fig.~\ref{figure:ratio}). The reason for this is not entirely clear but it may be that the mass of the \htvao\ emitting gas in reality is slightly larger than the values estimated from this work.
 The wind mass-loss rates are estimated to be \about \expo{-6}~\msunyr\ for L\,1448 and L\,1157, and one order of magnitude lower for \vlarhooph. Assuming that the luminosity of the central source\footnote{$L_{\rm{bol}}$~=~8.3, 8.4, and $<$2.0 for L\,1448, L\,1157 and \vlarhooph\ respectively \citep{Froebrich:2005lr}.} is entirely due to accretion and adopting a radius to mass ratio of 5, this leads to similar values for the mass-accretion rates \citep[see e.g.][]{Stahler:1980kx}.

\vspace{-0.2cm}\section{Conclusions}
\label{section:conclusions}
The inferred values for power agrees to within a factor of 3, when using the different molecular tracers. Similarly, the estimated values for thrust are equal to within a factor of 4. This may indicate that the ejection mechanism that is responsible for the motion of the cold gas, is the same as the one that sets the warm gas into motion. It is, however, also possible that this is a characteristic of young outflows. This study also reveals that the emission from \htvao\ traces a gas component that presumably operates on the same time-scales as CO, i.e. \htvao\ is not detected at very different velocities compared to CO. The fact that the estimated values are the same (within a factor of a few) for all observed sources supports previous estimates based on CO observations. We also find it likely, that ground-based CO observations are adequate, when assessing the impact of outflows on their environment. 

\begin{acknowledgements}
P. Bjerkeli wish to thank Eva Wirstr\"om for interesting discussions.
 The authors appreciate the support from the Swedish National Space Board (SNSB). 
HIFI has been designed and built by a consortium of institutes and university departments from across Europe, Canada and the United States under the leadership of SRON Netherlands Institute for Space Research, Groningen, The Netherlands and with major contributions from Germany, France and the US. Consortium members are: Canada: CSA, U.Waterloo; France: CESR, LAB, LERMA, IRAM; Germany: KOSMA, MPIfR, MPS; Ireland, NUI Maynooth; Italy: ASI, IFSI-INAF, Osservatorio Astrofisico di Arcetri- INAF; Netherlands: SRON, TUD; Poland: CAMK, CBK; Spain: Observatorio Astronomico Nacional (IGN), Centro de Astrobiologia (CSIC-INTA). Sweden: Chalmers University of Technology - MC2, RSS \& GARD; Onsala Space Observatory; Swedish National Space Board, Stockholm University - Stockholm Observatory; Switzerland: ETH Zurich, FHNW; USA: Caltech, JPL, NHSC. 
\end{acknowledgements}
\bibliographystyle{aa} % style aa.bst
\bibliography{papers}

\end{document}